\documentclass[runningheads]{llncs}
\usepackage[T1]{fontenc}
\usepackage{graphicx}
\usepackage{graphicx}
\usepackage[T1]{fontenc}
\usepackage[utf8]{inputenc}
\usepackage{graphicx}
\usepackage{microtype}
\usepackage{cite}
\usepackage[hyphens]{url}
\usepackage{amssymb}
\usepackage{amsmath}
\usepackage[mathscr]{euscript}
\usepackage{booktabs}
\usepackage[misc]{ifsym}
\usepackage{multirow}
\usepackage{tabularx}
\usepackage{diagbox}
\usepackage{adjustbox}
\usepackage{tikz}
\usepackage{soul}
\usepackage{enumitem}
\usepackage{caption}
\usepackage{subcaption}
\usepackage{comment}
\usepackage{color, colortbl}
\usepackage{array}
\newcolumntype{J}{>{\centering\arraybackslash}m{12cm}}

\usepackage{array}
\newcolumntype{L}{>{\centering\arraybackslash}m{7.5cm}}

\usepackage{array}
\newcolumntype{j}{>{\centering\arraybackslash}m{4cm}}

\pdfoutput=1

\usepackage[colorlinks = false, allcolors=blue]{hyperref}

\usepackage{ragged2e}

\newcommand{\Nayoung}[1]{\textcolor{black}{{#1}}}
\newcommand{\DM}[1]{\textcolor{black}{{#1}}}
\newcommand{\David}[1]{\textcolor{black}{{#1}}}
\newcommand{\Lu}[1]{\textcolor{black}{{#1}}}

\definecolor{Gray}{gray}{0.9}

\begin{document}
%

\title{Bridging the Gap: Commonality and Differences between Online and Offline COVID-19 Data}


\titlerunning{Bridging the Gap}
%
\author{Nayoung Kim\inst{1}  \and
Ahmadreza Mosallanezhad\inst{1} \and
Lu Cheng\inst{1\and 2}  \and
Baoxin Li\inst{1}  \and
Huan Liu\inst{1} }
\authorrunning{Kim et al.}
%
\institute{Arizona State University, Tempe, AZ, USA \and University of Illinois Chicago, Chicago, Illinois, USA \\
\email{\{nkim48,amosalla,lcheng35,baoxin.li,huanliu\}@asu.edu}}
\maketitle              
\begin{abstract}
\David{With the onset of the COVID-19 pandemic, news outlets and social media have become central tools for disseminating and consuming information. Because of their ease of access, users seek COVID-19-related information from online social media (i.e., online news) and news outlets (i.e., offline news). Online and offline news are often connected, sharing common topics while each has unique, different topics. A gap between these two news sources can lead to misinformation propagation. For instance, according to the Guardian, most COVID-19 misinformation comes from users on social media. Without fact-checking social media news, misinformation can lead to health threats. In this paper, we focus on the novel problem of bridging the gap between online and offline data by monitoring their common and distinct topics generated over time. We employ Twitter (online) and local news (offline) data for a time span of two years. Using online matrix factorization, we analyze and study online and offline COVID-19-related data differences and commonalities. We design experiments to show how online and offline data are linked together and what trends they follow.} 



\end{abstract}
\section{Introduction}

\Lu{The COVID-19 pandemic has been accompanied by a massive infodemic involving a variety of topics, such as vaccination, inflation, and so on. Communication of important information during emergency situations is critical \cite{reynolds2002crisis,cheng2020tracking} for taking actions to contain disease. With its convenience, easy access, and large volume, news media becomes an expected means to a pandemic response. People seek and receive information from numerous sources such as newspapers, television, and increasingly, social media. For example, during the early stages of the COVID-19 pandemic, Twitter and news outlets such as Washington post reported about US states' social distancing mandates to help prevent the growth of the virus.}


\Lu{Due to the different nature of online social media data and offline news articles, information gaps may exist among online social media users and offline news media readers. These users and readers consist of scientists, medical and public health professionals, and the general public. A large information gap could contribute to the spread of misinformation in news media. Therefore, to better understand the relations between online social media data and offline news articles, we aim to monitor their common and distinct topics generated over time. In particular, connecting offline with online data can be beneficial in two main ways: \David{(1) Researchers can measure public interest over time and identify new themes of COVID-19-related discussion emerging quickly through data streams. (2) By providing users with useful information about COVID-19, it reduces the chance of exposing them to misinformation, i.e., a gap between offline and online topics can alert users about possible misinformation propagation. } }

\begin{figure}[t!]
    \centering
    \includegraphics[width=0.8 \textwidth]{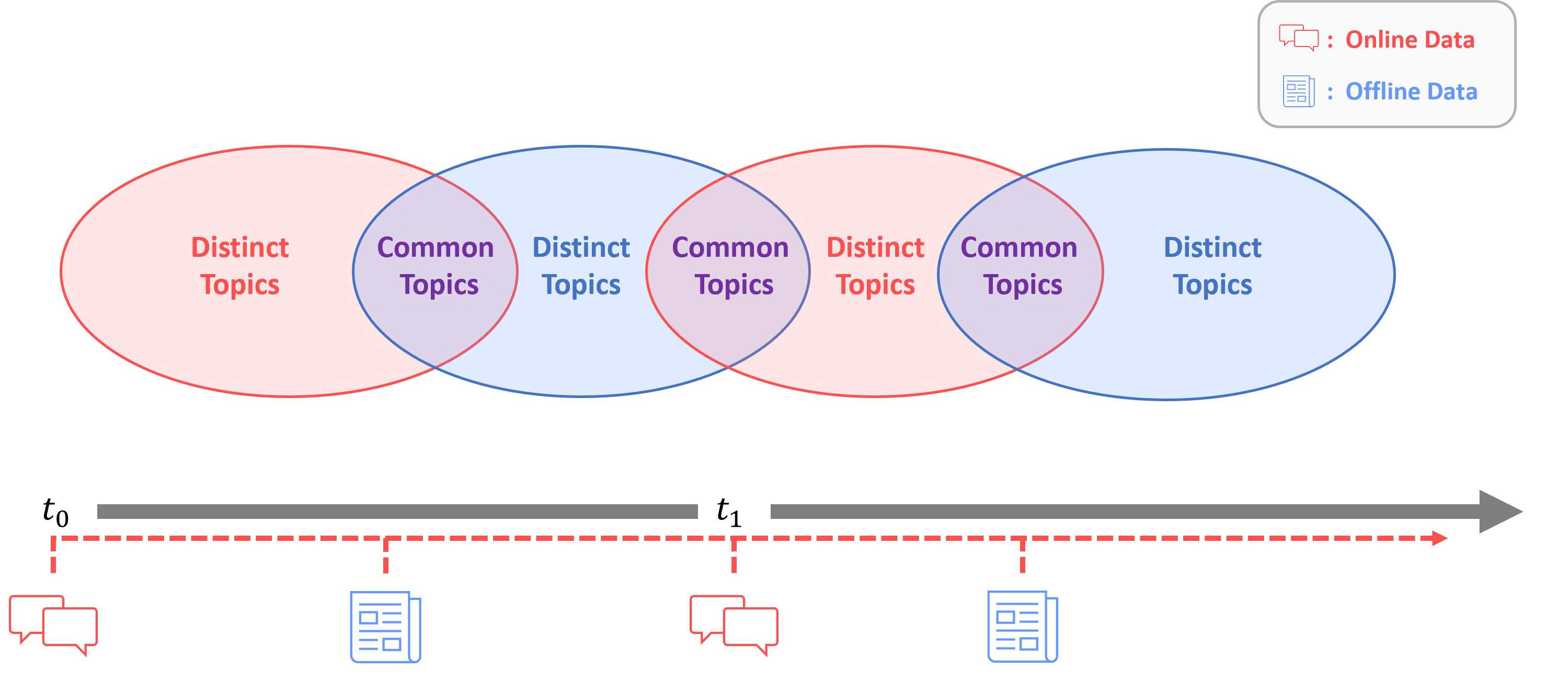}
    \caption{\Lu{Illustration of the common and distinct topics identified within two types of datasets: \textit{offline} and \textit{online}.}} 
    \label{fig:intro}
\end{figure}

\autoref{fig:intro} shows the problem setting. Our goal is to discover and link topics from the traditional and social media news sources, to understand their association. We study this problem on a COVID-19-related dataset~\cite{jiang2022covaxnet} collected through Twitter and Google News. This problem presents the following challenges: (1) The collected dataset contains a significant amount of noisy and irrelevant data. 
In the online data, not all the tweets follow a specific topic or provide useful information about an actual event. Consequently, the integration approach should discard noise and irrelevant data. 
(2) Identifying topics across two different data sources in a separate and joint setting. \Nayoung{The model should be able to find the association between them by leveraging the dataset properties.} 
Typically, Twitter data are short texts, including hashtags, user mentions, and covering any topics. In contrast, news articles published on media websites are long and structured.

\DM{To address the challenges above, we use a variant of the online Non-negative matrix factorization (NMF) method, joint ONMF. NMF methods are capable of dimensionality reduction and data clustering simultaneously. Because of our data's nature and its streaming property, we cannot use vanilla NMF. Further, the online NMF method cannot model discovered topics over time. As a result, we adopt joint ONMF to find common and distinct topics over time and link online and offline data. Specifically, we overcome the challenges by making the following contributions: (1) We study the novel problem of linking data related to COVID-19 from online social media and offline news sources, (2) We use joint ONMF to identify common and distinct topics across time, and (3) We conduct extensive experiments on real-world datasets to draw conclusions.}

\section{Related Work}

There have been a growing number of research on social media discourse associated with disasters or crises. For example, Martin et al.~\cite{martin2017leveraging} leveraged the spatiotemporal features of tweets to assess the responses on Hurricane Matthew, and Yeo et al.~\cite{yeo2020disaster} analyzed Twitter communications on 2016 Southern Louisiana flood recovery process.
Cheng et al.~\cite{cheng2020tracking} studied the problem of tracking disaster events through streaming data from Twitter. They proposed an online matrix factorization-based method to identify topics across time periods. 

As people share their emotions~\cite{lwin2020global} and opinions~\cite{cotfas2021longest} related to the COVID-19 pandemic on social media (e.g. Twitter, Facebook), a number of studies have also been conducted to explore the use of social media during COVID-19 epidemic. Although numerous researches have focused on online discourses~\cite{wicke2021covid, glandt2021stance, xiong2021social}, a few recent studies have tried to disclose the impact of various offline data delivered by traditional news sources. Crupi et al.~\cite{crupi2022echoes} showed the change of topics regarding COVID-19 vaccination in Italian user communities and the polarization of vaccine stances with common and distinct topics. Another work by Poddar et al.~\cite{poddar2022winds} identified distinct topics in online data and investigated how vaccine-related discourse has changed between pre and during COVID times. A few studies have utilized geo-tagged data on social media to map online data to real-world events and statistics~\cite{shen2020using, feng2021integrating}. Others tried to uncover the association between vaccine hesitancy and online user behavior~\cite{wilson2020social,macdonald2015vaccine} or online misinformation~\cite{lyu2022misinformation, pierri2022online}.

Our work is different than previous studies as we study the novel problem of linking COVID-19-related offline and online data. The aim of this problem is to examine how common and distinct topics around offline and online data change over time, and how offline events affect online data.




\section{Proposed Method}
In this section, we first review the basic NMF model and its online variant which is a modified matrix factorization for temporal data. Then, we discuss the main components of the proposed method in detail. Inspired by Cheng et al.~\cite{cheng2020tracking}, we employ the joint ONMF algorithm to extract the latent factors of streaming data using matrix factorization. \David{Considering a time span of 24 months, in every one-month time period, historical data is combined with newly arriving data to identify common and distinct topics. We perform the joint ONMF on COVID-19 tweets and its related news data to extract the topics.}


\subsection{Non-negative Matrix Factorization (NMF)}
NMF~\cite{lee1999learning} is a popular non-negative matrix decomposition algorithm which is widely used for analysis of multivariate data such as latent representation of text. Given a non-negative data matrix $\mathbf{V}$, NMF factorizes $\mathbf{V}$ into two lower rank non-negative matrices $\mathbf{W}$ and $\mathbf{H}$ as $\mathbf{V}\approx \mathbf{W} \mathbf{H}$.
Matrix $\mathbf{V} \in \mathbb{R}^{m\times n}$ indicates $m$ data, each having $n$ attributes. Matrix $\mathbf{H}$ is a feature matrix with each row represents a latent factor. $\mathbf{W} \in \mathbb{R}^{m\times k}$ is a coefficient matrix, reflecting the association weights between the data samples and the extracted latent factors. NMF is solved by minimizing the distance between the original matrix $\mathbf{V}$ and the reconstructed matrix $\mathbf{V'} = \mathbf{W} \mathbf{H}$.


\David{The NMF method assumes both latent factors and input data are static and do not change over time. Thus, we cannot directly apply this method for our problem. One simple solution to apply NMF on streaming data is to feed NMF with updated data matrix as new subset of the data joins at each time step $t$. However, this approach underutilizes the previous NMF results in the upcoming factorization results that is calculated for the newly collected data~\cite{cao2007detect}. Cao et al.~\cite{cao2007detect} proposes Online-NMF (ONMF) to decompose matrix $\mathbf{V}_{t+1}$ based on a new batch of data $\mathbf{U}$ at time step $t+1$ and the previous matrix $\mathbf{V}_t$. Formally, the ONMF problem is defined as $\mathbf{V}_{t+1}= \begin{pmatrix} \mathbf{V}_{t}\\ \mathbf{U} \end{pmatrix}  \approx \mathbf{W}_{t} \mathbf{H}_{t}$, which is solved by minimizing the mean squared error between $\mathbf{V}$ and $\mathbf{WH}$~\cite{cao2007detect}. 
}

\subsection{Linking Offline and Online Data}
ONMF can efficiently update the latent factors (i.e., topics) in streaming data. However, it is hard to see the global semantic changes in the generated topics. To solve this problem, inspired by \cite{cheng2020tracking}, we use a modified ONMF-based approach, joint ONMF, to discover commonness and distinctiveness in topics. Assuming there are $k$ topics in the given documents, joint ONMF identifies $k_c$ and $k_d = k - k_c$ which are defined as the number of \textit{common topics} and \textit{distinct topics}, respectively. To alleviate the massive computational cost proportional to the growing data matrix $\mathbf{V}$, this method uses one of the decomposed low rank matrices $\mathbf{H}_t$ to derive topics. Since $\mathbf{H}_t$ is fixed at time step $t$, we perform the linear transformation $\mathbf{H}^*\approx \mathbf{L}^*\mathbf{H}_t$ to dynamically tune the dependency between $\mathbf{H}_t$ and newly arrived data $\mathbf{U}$.
In this transformation, $(\mathbf{L}^*\in\mathbb{R}^{k \times k})$ is a low rank matrix that is used to adjust the dependency between $\mathbf{H}_t$ and $\mathbf{U}$. 

Joint ONMF aims to minimize the distance between common topics in $\mathbf{H}^*$ and $\mathbf{H}_U$ and maximize the distance between distinct topics in $\mathbf{H}^*$ and $\mathbf{H}_U$.
Finally, we use the following objective function to find common and distinctive topics:
\begin{align} \label{eq:optim} \nonumber
\min_{\mathbf{W}_U, \mathbf{H}_U, \mathbf{H}^*, \mathbf{L}^*} \frac{1}{2}\|\mathbf{H}^* - \mathbf{L}^* \mathbf{H}_t \|^2_F +& \frac{1}{2}\|\mathbf{U} - \mathbf{W}_{U_c} \mathbf{H}_{U_c} - \mathbf{W}_{U_d} \mathbf{H}_{U_d} \|^2_F \\ 
+& \alpha f_c(\mathbf{H}^*_c, \mathbf{H}_{U_c}) + \beta f_d(\mathbf{H}^*_d, \mathbf{H}_{U_d}),
\end{align}
where the second term denotes the reconstruction error on matrix $\mathbf{U}$, which is factorized into $\mathbf{W}_U$ and $\mathbf{H}_U$ $(\mathbf{U} \approx \mathbf{W}_U \mathbf{H}_U)$. $f_c$ and $f_d$ are the commonness score and distinctiveness score described in \autoref{section:exp}. We leverage parameters $\alpha$ and $\beta$ to find the optimal balance between three different \Nayoung{objectives:}
the accuracy of matrix reconstruction, common, and distinct topics.

This method identifies the common and distinct topics over time within one dataset. To examine the relationship between the datasets with different sources, we apply the method on both datasets simultaneously. Specifically, we split each dataset by months as delimiters and concatenate the subsets of tweets and news pieces one after another along the timeline. (e.g., Tweets of January 2020 $||$ News of January 2020 $||$ Tweets of February 2020 $||$ News of February 2020 $\cdots$). Starting with the Tweets of January 2020, the model outputs the common topics between each month (i.e., topics that appear both in $t$ and $t+1$, considering $t$ as month) and distinct topics (i.e., two unique topics at time before $t$ and time $t+1$) whenever new data arrives at every time stamp $t$. The result distinct topics could give an insight on the main theme and statements raised within a group of people on Twitter, along with the important events recorded by the press happened in the specific time period. On the contrary, the output of common topics implies the tweets' topics are closely related to the news pieces. 

\section{Experiments}\label{section:exp}
We conduct qualitative and quantitative evaluation on the performance of the joint ONMF model for extracting topics and gaining insights from both online and offline data. We aim to answer two research questions: (\textbf{Q1}) How well joint ONMF extracts common and distinct topics across online and offline data?
(\textbf{Q2}) How does the offline data impact the online data?
To answer \textbf{Q1}, we compare the performance of joint ONMF with baselines based on various evaluation metrics. To answer \textbf{Q2}, we design experiments to measure the distinctiveness of online and offline data. 

\subsection{Experimental Design}
We use the CoVaxNet dataset\footnote{\url{https://github.com/jiangbohan/CoVaxNet}} that includes $4,790$ offline news articles and $1,831,220$ online tweets within a time span of two years from January 2020 till December 2021~\cite{jiang2022covaxnet}. 
\Nayoung{We further use Newspaper3k\footnote{\url{https://github.com/codelucas/newspaper}} to extract \textit{title}, \textit{body}, and \textit{summary} from news pieces.} The online data is first filtered to select $20K$ representative random tweets. Finally, we extract the initial textual features using TF-IDF scores from both datasets. A large TF-IDF value indicates that the term can \Nayoung{better} distinguishes documents from each other.

To design experiments, we start with online data from January 2020 followed by offline data from the same time period. Assuming the topics of tweets vary in accordance with newly uploaded news pieces within certain period, we assign tweets and news by turns \Nayoung{monthly}.
We choose $k_c = 2$ and $k_d = 3$. The optimization parameters $\alpha$ and $\beta$ of \autoref{eq:optim} are set to $1,000$ and $0.1$, respectively. Finally, the step size of the joint ONMF method is set to $100$.

To evaluate the effectiveness of joint ONMF, we consider the three baselines. \textbf{Standard NMF (SNMF)}~\cite{lee2000algorithms} approach repeatably calculates the latent factors using the entire dataset whenever a new batch of data is received. \textbf{ONMF}~\cite{cao2007detect} method is a modified variant of NMF that focuses on handling large-scale and streaming data. Finally, \textbf{Pseudo-Deflation}~\cite{kim2015simultaneous} approach uses a variant of ONMF, to identify common and distinct topics between two documents. \Nayoung{For fair comparison between joint ONMF and the baselines, we evaluate the retrieved topics within the same time period and hyperparameters.}

\begin{figure*}[t!]\vspace{-10pt}
     \centering
     \begin{subfigure}[b]{0.40\textwidth}
         \centering
         \includegraphics[width=\textwidth]{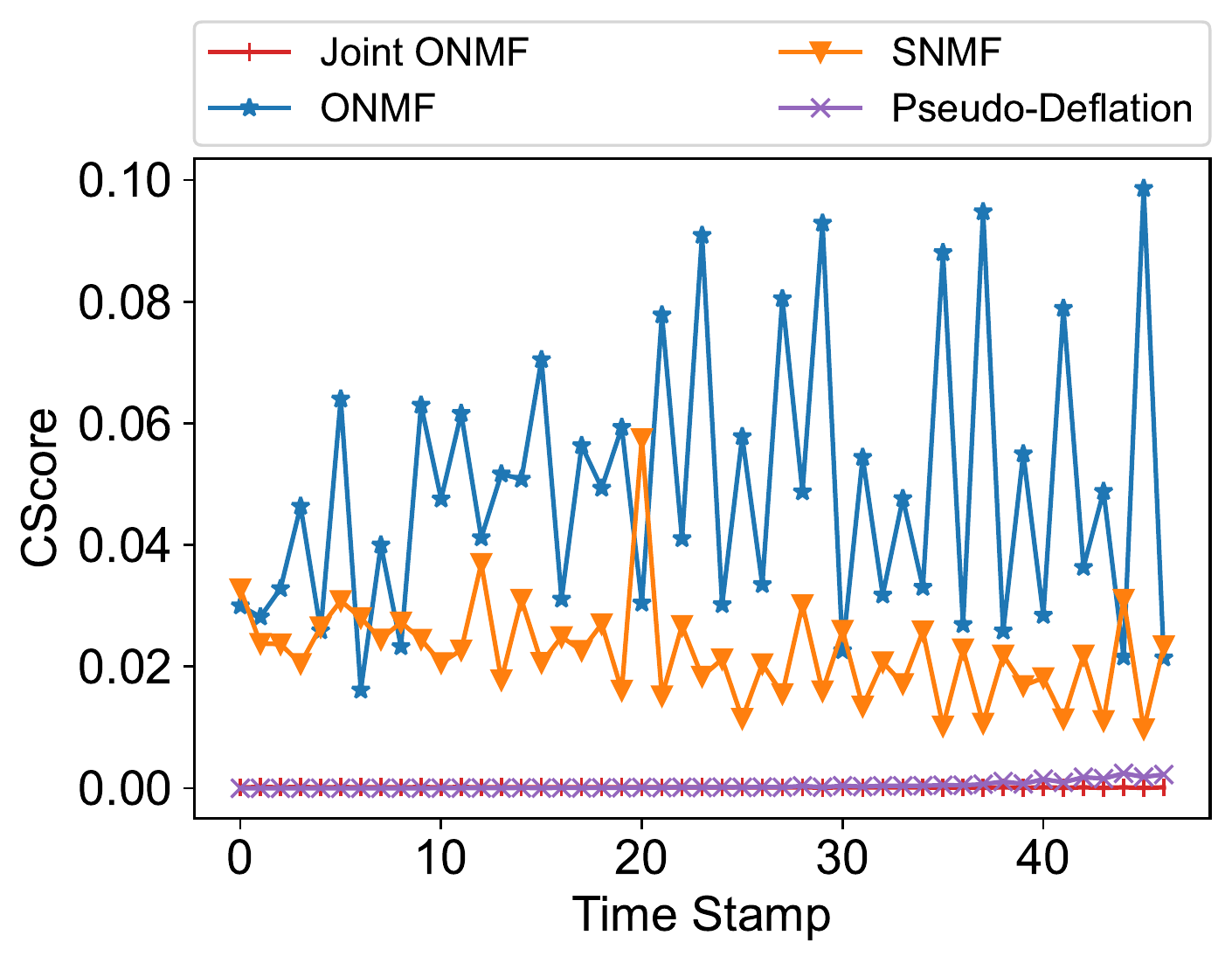}
         \caption{CScore}
         \label{fig:common23}
     \end{subfigure}
     \begin{subfigure}[b]{0.40\textwidth}
         \centering
         \includegraphics[width=\textwidth]{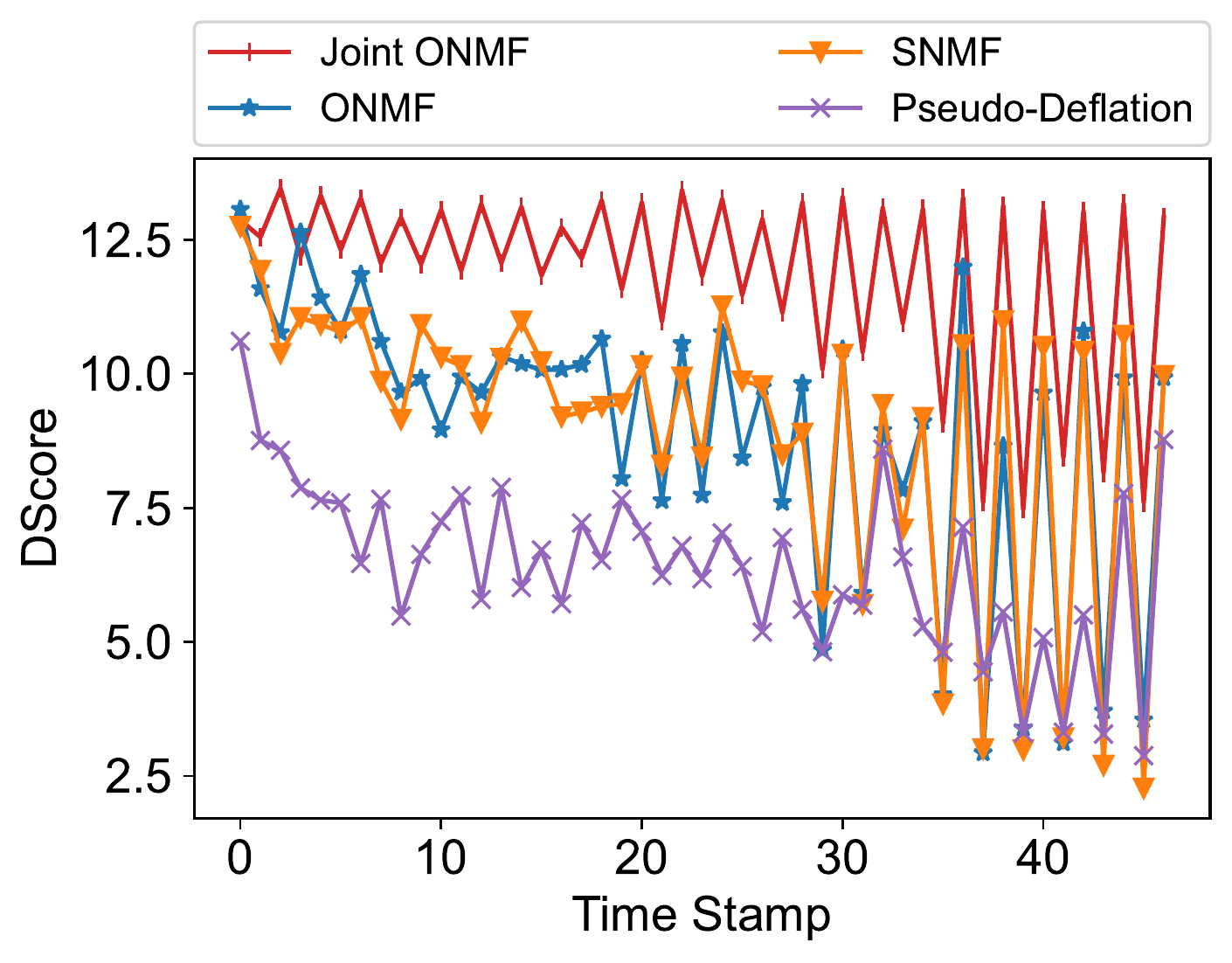}
         \caption{DScore}
         \label{fig:differ23}
     \end{subfigure}
     \vskip\baselineskip
     \vspace{-4mm}
     \begin{subfigure}[b]{0.40\textwidth}
         \centering
         \includegraphics[width=\textwidth]{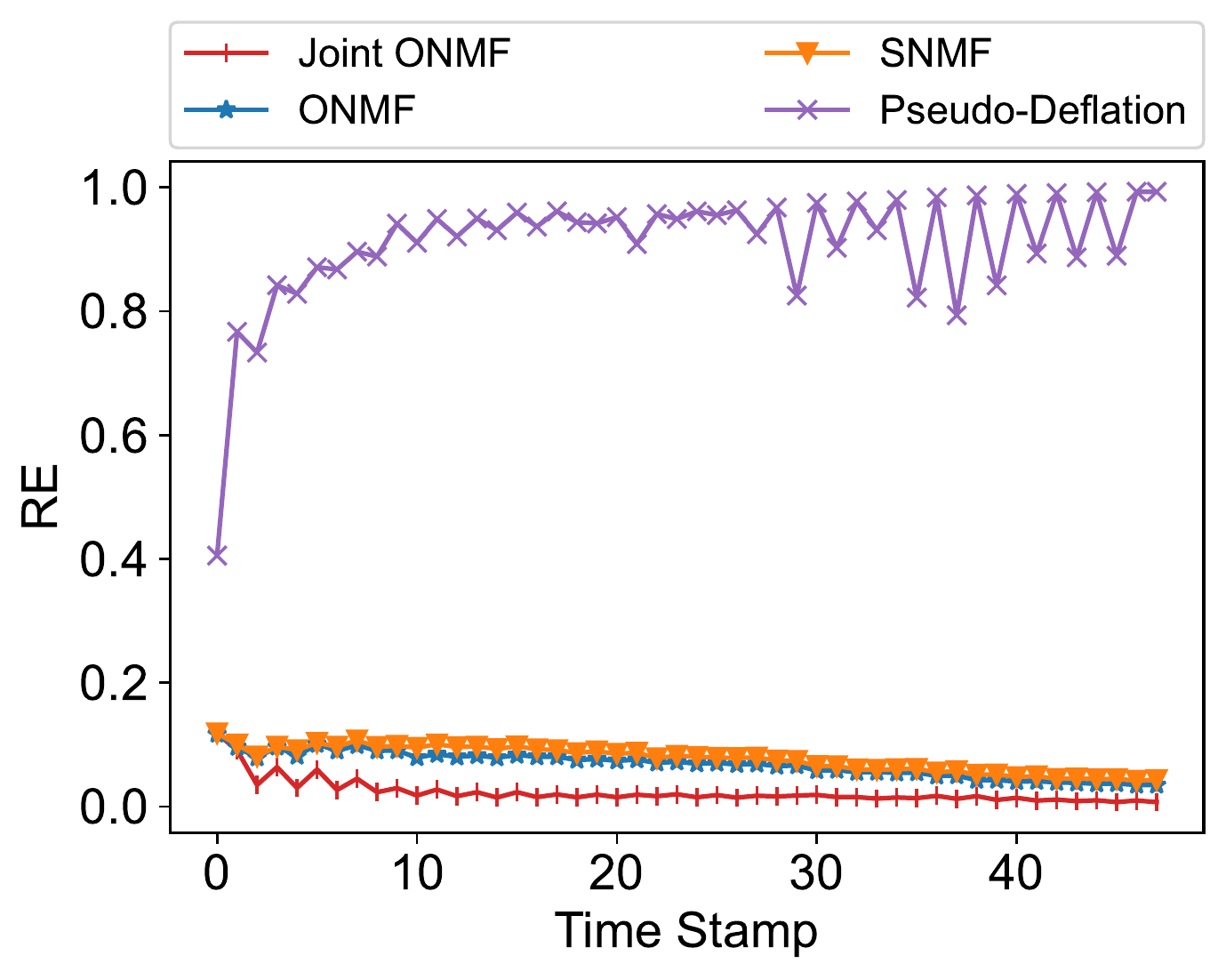}
         \caption{RE}
         \label{fig:error_log23}
     \end{subfigure}
     \begin{subfigure}[b]{0.40\textwidth}
         \centering
         \includegraphics[width=\textwidth]{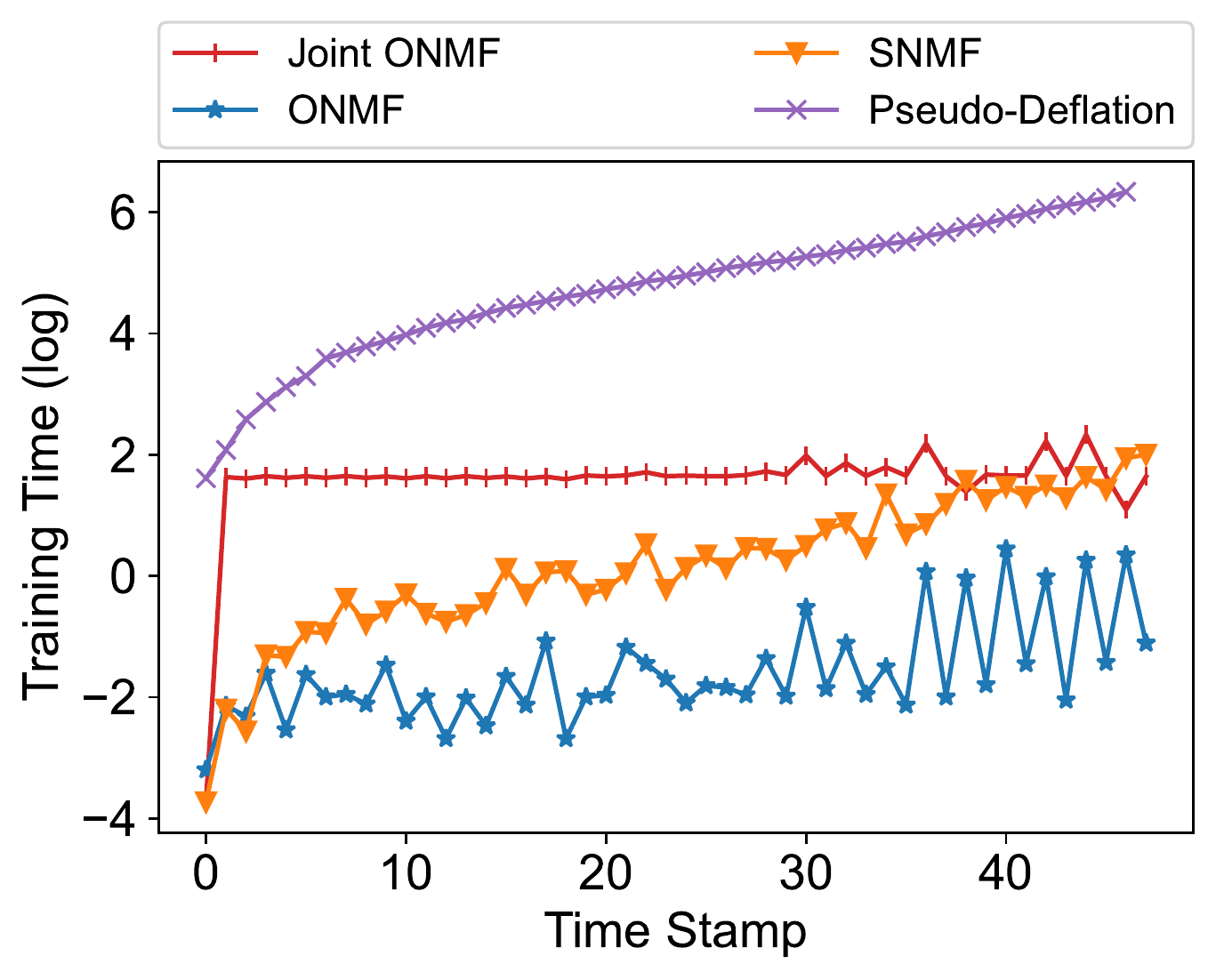}
         \caption{Training Time}
         \label{fig:Time_log23}
     \end{subfigure}
     
        \caption{Performance comparison of different methods}
        \label{fig:four graphs}
\end{figure*}

\subsection{Experimental Results}
Following Kim et al.~\cite{kim2015simultaneous}, we evaluate the performance of joint ONMF using commonness score, distinctiveness score, and reconstruction error: %

\noindent \textbf{Commonness Score (CScore)} stands for the similarity between the $k$ common topics at time $t$ and $t+1$ for every time stamp, calculated as $\text{CScore} = \frac{1}{k_c}\|\mathbf{H}^*_c - \mathbf{H}_{\mathbf{U}_c} \|^2_F$. A smaller CScore indicates better efficiency of the method on extracting common topics.
\noindent \textbf{Difference Score (DScore)} uses average symmetric Kullback-Leibler (KL) divergence between all the distinct topic pairs. A large DScore indicates the distribution of the obtained distinct topics are discrete.
\begin{align}\nonumber
\text{DScore} = \frac{1}{2k^2_d}\sum^{k_d}_{i}\sum^{k_d}_{j}[&h^{*i}_d \log(h^i_d)^T + h^i_{\mathbf{U}_d} \log(h^i_{\mathbf{U}_d})^T - \\ 
&- h^{*i}_d \log(h^j_d)^T - h^j_{\mathbf{U}_d} \log(h^{*i}_{\mathbf{U}_d})^T] 
\end{align}
\noindent \textbf{Reconstruction Error (RE)} estimates the loss of the NMF on the newly arriving data $\mathbf{U}$ at each time stamp $t$ based on 
$\min_{\mathbf{W}, \mathbf{H}} \|\mathbf{V} -  \mathbf{W} \mathbf{H}\|$. Smaller RE indicates the initial data matrix $\mathbf{U}$ can be reconstructed accurately. 

\begin{figure*}[t!]\vspace{-6pt}
     \centering
     \begin{subfigure}[b]{0.34\textwidth}
         \centering
         \includegraphics[width=\textwidth]{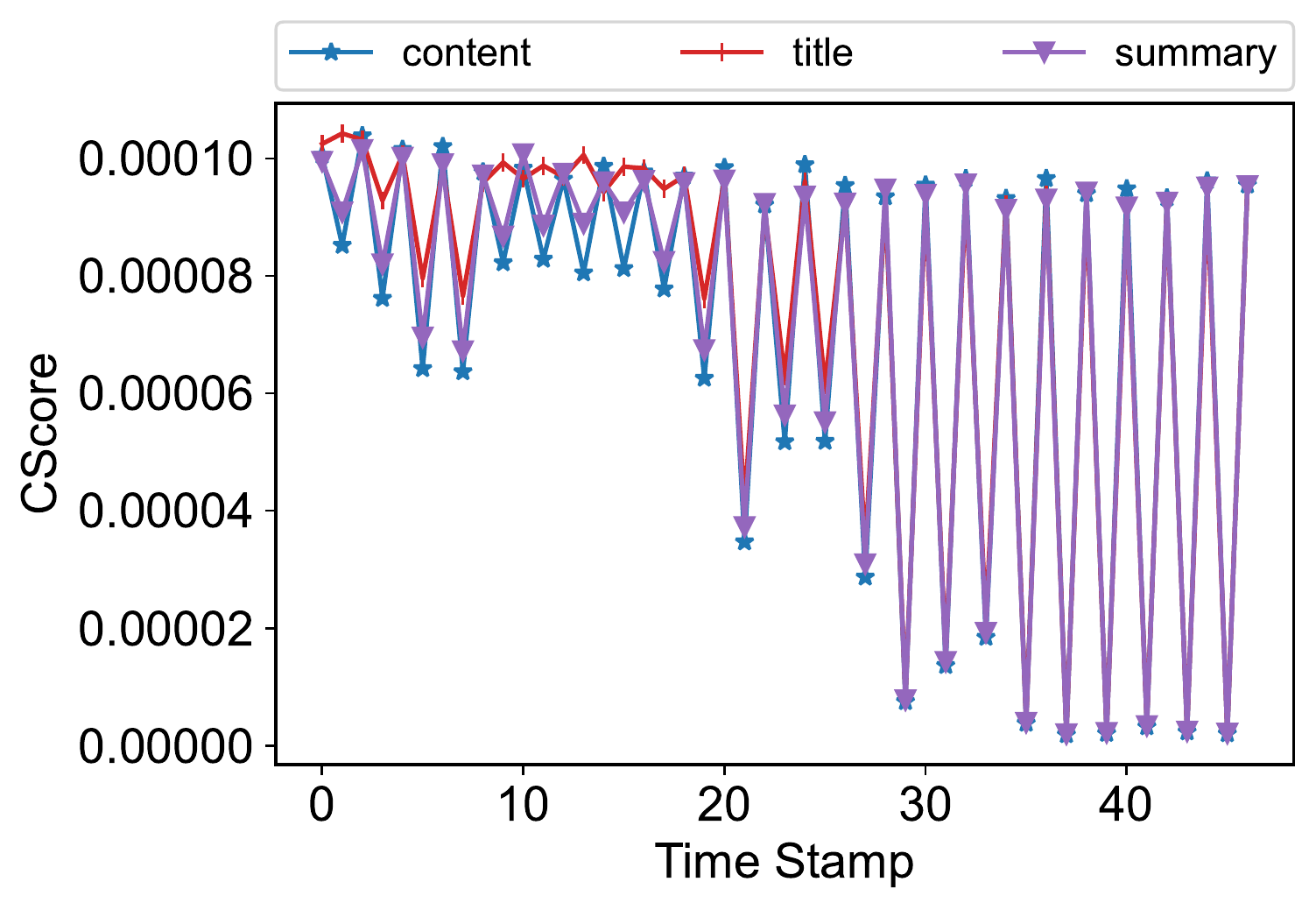}
         \caption{CScore}
         \label{fig:TDF_C}
     \end{subfigure}
     \begin{subfigure}[b]{0.31\textwidth}
         \centering
         \includegraphics[width=\textwidth]{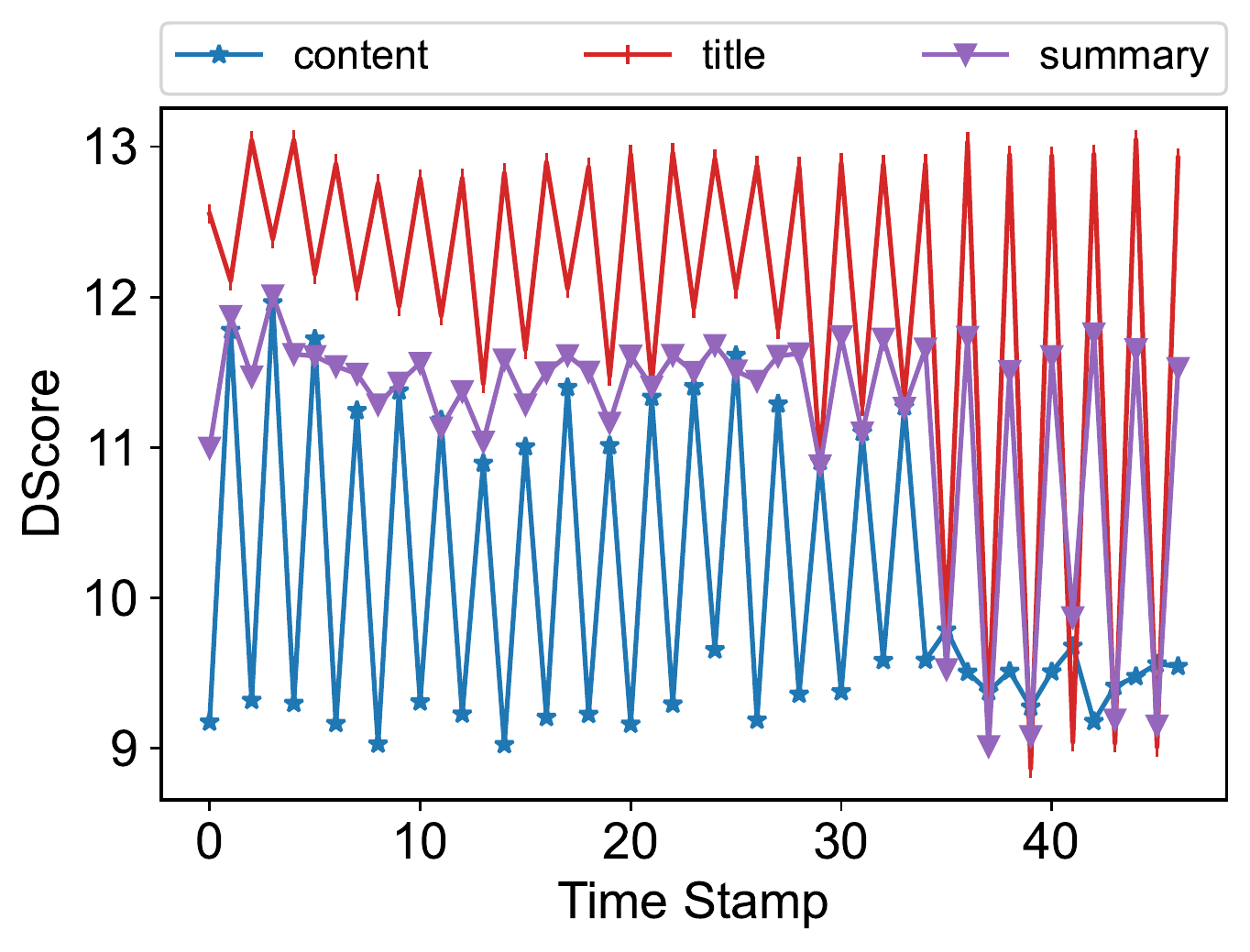}
         \caption{DScore}
         \label{fig:TDF_D}
     \end{subfigure}
     \begin{subfigure}[b]{0.31\textwidth}
         \centering
         \includegraphics[width=\textwidth]{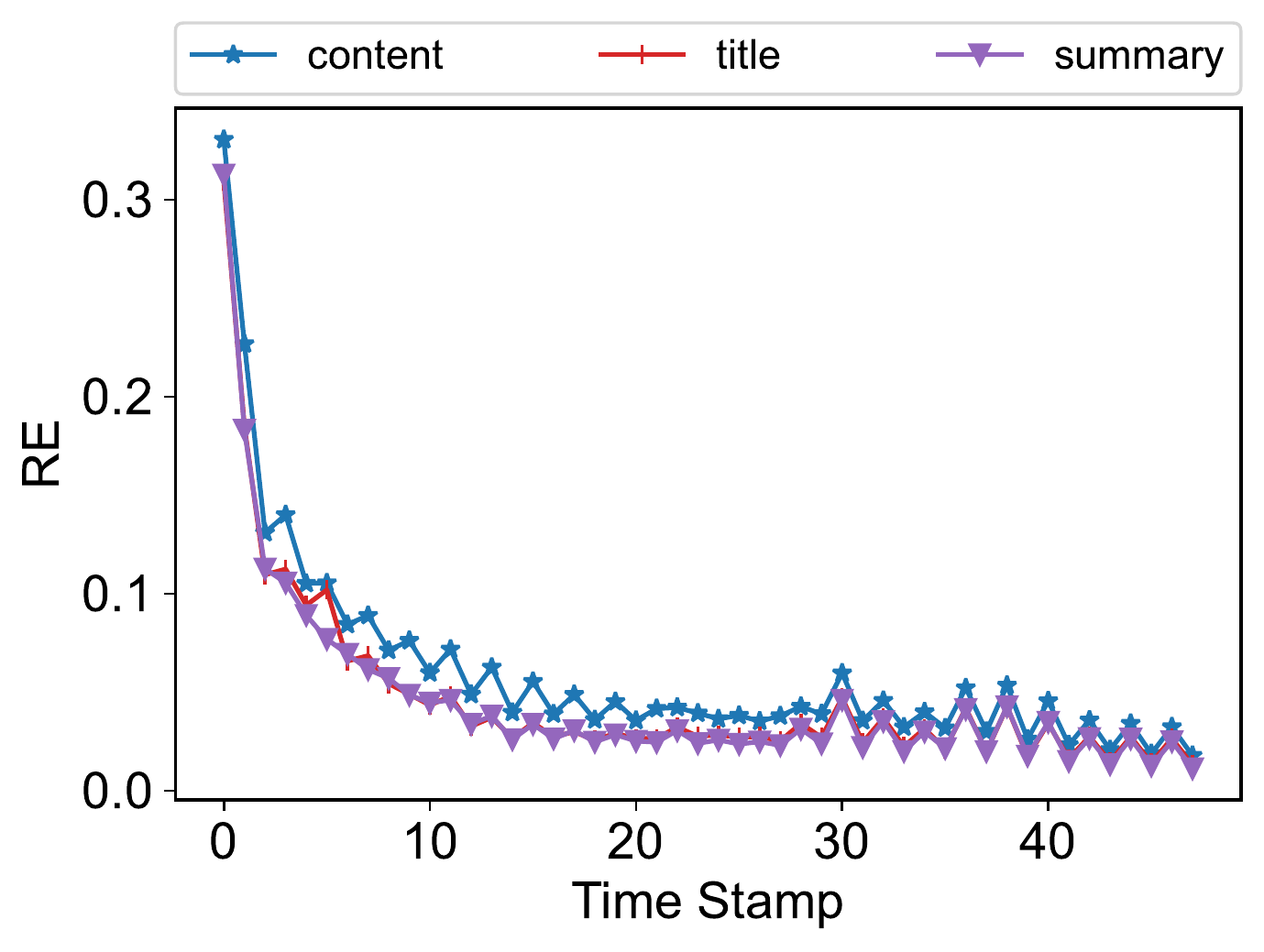}
         \caption{RE}
         \label{fig:TDF_E}
     \end{subfigure}
     
    \caption{Performance comparison with different data metrics of offline data}
    \label{fig:three graphs}
\end{figure*}

We answer the two questions \textbf{Q1} and \textbf{Q2} using quantitative and qualitative experiments to evaluate the quality of common and distinct topics in the integrated dataset, and study the effect of offline data on the online data.

\vspace{5pt}\noindent \textbf{Quantitative Results:}\vspace{5pt}

\noindent\textbf{Performance (Q1).} To calculate CScore and DScore, we directly apply them on \textsc{joint ONMF} and \textsc{Pseudo-Deflation} as they explicitly state which topics are the common and distinct. For other baseline models we assign topics with the smallest CScore as common topics $k_c$, while the others are treated as distinct ones $k_d$. \autoref{fig:four graphs} illustrates the evaluation results including computation time of each method. From these results, we conclude that the \textsc{joint ONMF} method outperforms \textsc{ONMF} and \textsc{SNMF} with a large gap in CScore and DScore. \textsc{Pseudo-Deflation} has comparable CScore, however, the DScore is significantly lower than \textsc{joint ONMF}. DScores of all four methods fluctuates severely in later time periods due to the noisy and unbalanced nature of the dataset. 
Among the baselines, \textsc{Pseudo-Deflation} presents remarkably smaller RE. Besides \textsc{Pseudo-Deflation}, \textsc{SNMF} can often obtain the best RE because the goal of \textsc{SNMF} focuses on minimizing RE while other online topic models seek to balance between RE and computational efficiency. The \textsc{joint ONMF} method presents competitive RE compared to \textsc{SNMF} and \textsc{ONMF} as it separately conducts matrix factorization on data matrix $\mathbf{U}$.

Considering all three evaluation metrics and the computational cost, the \textsc{joint ONMF} approach attains the most optimal balance among the baselines. For instance, it can have the smallest CScore and the largest DScore, meanwhile achieving negligible reconstruction error similar to \textsc{SNMF}. \autoref{fig:Time_log23} shows that \textsc{joint ONMF} is slightly slower than two baselines yet maintains consistent training time, making it comparatively computationally efficient. 

\noindent\textbf{Impact (Q2).} According to \autoref{fig:three graphs}, one interesting observation is that the amount of offline information inconsistently influence the performance of \textsc{joint ONMF} over each measurements. As mentioned earlier, we measure the performance on \{`title', `summary', `body'\} of each news piece, assuming that `title' text has the smallest, and the `body' text has the largest quantity of information. Results show that given more information, the model can better reconstruct original data but worse in distinguishing distinct topics.

\begin{table*}[t!]\vspace{-17pt}
    \centering
    \caption{Examples of topics from Jan 1st to Mar 31st, 2020. $D_T$ and $D_N$ show distinct topics of tweet and news title, respectively. $C$ shows the common topics of between offline and online data.}
    \begin{tabularx}{\textwidth}{X|X|X|X|X|X|X|X|X}
        \hline
        \multicolumn{3}{c|}{\textbf{Jan 2020}} & \multicolumn{3}{c|}{\textbf{Feb 2020}} &
        \multicolumn{3}{c}{\textbf{Mar 2020}} \\

        \hline
        \scriptsize\textbf{$D_T$} & \scriptsize\textbf{$C$}  & \scriptsize\textbf{$D_N$} & \scriptsize\textbf{$D_T$} & \scriptsize\textbf{$C$} & \scriptsize\textbf{$D_N$} & \scriptsize\textbf{$D_T$} & \scriptsize\textbf{$C$} & \scriptsize\textbf{$D_N$} \\
        \hline\hline
        \rowcolor{Gray}
        \scriptsize\upshape{misinfor- mation} & \scriptsize\upshape{oil} & \scriptsize\upshape{antigen} & \scriptsize\upshape{deadly} & \scriptsize\upshape{flu} & \scriptsize\upshape{{\fontsize{5pt}{5pt}\selectfont Remdesivir}} & \scriptsize\upshape{misinfor- mation} & \scriptsize\upshape{patent} & \scriptsize\upshape{immunity} \\

        \scriptsize\upshape{children} & \scriptsize\upshape{Japan} & \scriptsize\upshape{China} & \scriptsize\upshape{fund} & \scriptsize\upshape{{\fontsize{5pt}{5pt}\selectfont antivaccine movement}} & \scriptsize\upshape{influenza season} & \scriptsize\upshape{worry} & \scriptsize\upshape{mask wear} & \scriptsize\upshape{Donald Trump} \\     
        
        \rowcolor{Gray}
        \scriptsize\upshape{thank} & \scriptsize\upshape{religious exemption} & \scriptsize\upshape{HPV cancer} & \scriptsize\upshape{happy} & \scriptsize\upshape{pediatric} & \scriptsize\upshape{South Korea} & \scriptsize\upshape{child killer} & \scriptsize\upshape{candidates} & \scriptsize\upshape{CDC} \\     

        \scriptsize\upshape{kill} & \scriptsize\upshape{Ebola} & \scriptsize\upshape{Donald Trump} & \scriptsize\upshape{thank} & \scriptsize\upshape{conspiracy} & \scriptsize\upshape{afford insurance} & \scriptsize\upshape{{\fontsize{5pt}{5pt}\selectfont propaganda}} & \scriptsize\upshape{INOVIO (Pharmaceuticals)} & \scriptsize\upshape{adjuvants} \\ 
        
        \rowcolor{Gray}
        \scriptsize\upshape{amazing} & \scriptsize\upshape{Johnson \& Johnson} & \scriptsize\upshape{tuber- culosis} & \scriptsize\upshape{autism} & \scriptsize\upshape{{\fontsize{5pt}{5pt}\selectfont Pneumonia}} & \scriptsize\upshape{peptid} & \scriptsize\upshape{prevent disease} & \scriptsize\upshape{egg-based vaccine} & \scriptsize\upshape{Summit} \\
        
        \hline
    \end{tabularx}
    \label{tab:examples}
\end{table*}

\vspace{3pt}\noindent \textbf{Qualitative Study:}\vspace{3pt}

\noindent\textbf{Performance (Q1).} We perform an extensive qualitative study on the validity of the identified topics. \autoref{tab:examples} shows top five ranked topics discovered by the \textsc{joint ONMF} method on the integrated online and offline data during the first three time periods of the pandemic. The extracted common topics are closely related to temporal and/or newly rising events (e.g., \upshape{Ebola}, \upshape{Remdesivir}, \upshape{egg-based vaccine}) which are compelling and provocative as well (i.e., topics discussed online change over time and selective real-world events). 
Meanwhile, distinct topics consist of specific keywords relevant to COVID-19 or medical events. Note that according to \autoref{tab:examples}, exceptionally large portion of emotional words account for the common topics of the online data. 

\noindent\textbf{Impact (Q2).} We take a deeper analysis on the common topics to observe how the two datasets are linked. For example, in \autoref{tab:casestudy}, we sample tweets and news headlines containing the topic keyword \textit{oil}. It is worth mentioning that the users not only talk about the facts and opinions, but also they expose their strong emotions (e.g., {\footnotesize\fontfamily{cmtt}\selectfont i will laugh, lol}) and groundless claims (e.g., {\footnotesize\fontfamily{cmtt}\selectfont oil companies just hid their own research data}) that could falsely lead to misinformation. We could find similar characteristics in multiple other tweet samples. This implies that the offline data (i.e., local events and news) play a substantial role in the shift of online behavior and potential biases of users.

\begin{table*}[t!]\vspace{-15pt}
    \centering
    \caption{Examples of different sources from common topic }
    \begin{tabularx}{\textwidth}{c|c|m{28em}}
        \hline
        \textbf{Topic} & \textbf{Source} & \textbf{Examples} \\\hline\hline
        
        \multirow{9}{*}{\textit{oil}} & \multirow{1}{*}{News Articles} & 
        - Russia halts oil to Belarus, but transit to Europe still flowing (Jan 3, 2020)\\ 
        \cline{2-3}
        
        & \multirow{7}{*}{Tweets} & - Lol oh yes definitely max single handedly created the need smh he is just one of the ... antivaxxers of the oil world (Jan 4, 2020) \\ 
        & & - The oil companies just hid their own research data now all the sceptics have left like flat earthers ... (Jan 5, 2020) \\
        & & - I will laugh at anyone who sneers at China for this given that America is a rapidly decaying shithole filled with ... climate denialism antivaxxers and oil companies that own multiple politicians (Jan 31, 2020) \\\hline\hline
        
        \multirow{5}{*}{\textit{trial}} & \multirow{1}{*}{News Articles} & 
        - Coronavirus Vaccine: NIH Goes Straight to Human Trials (Mar 16, 2020) \\\cline{2-3}
        
        & \multirow{3}{*}{Tweets} & - Let’s go science NIH clinical trial of investigational vaccine for covid 19 begins ... (Mar 16, 2020) \\  
        & & - Why i am volunteering to get the fastest startup of a vaccine trial ever ... (Mar 31, 2020)
        \\\hline\hline
        
        \multirow{11}{*}{\textit{delta} } & \multirow{3}{*}{News Articles} & 
        - Pfizer says COVID vaccine is highly effective against Delta variant (Jun 24, 2021)\\ 
        && - Moderna Says Studies Show Its Vaccine Is Effective Against The Delta Variant (Jun 30, 2021)
        \\\cline{2-3}
        
        & \multirow{5}{*}{Tweets} & - This is what I've been saying the delta variant is growing and now there is a new lamda variant while we can't live in fear over every variant that pops up we will still need to be cautious until we learn what they do and whom they affect (July 8, 2021) \\ 
        & & - Get vaccinated or get sick but do not punish those who are vaccinated I am fed up with restrictions on us because of and citing rise of delta variant Los Angeles reports 165 percent increase in covid cases (July 9, 2021) \\\hline
        
    \end{tabularx}
    \label{tab:casestudy}
\end{table*}

\vspace{-5pt}\section{Conclusion and Future Work}\vspace{-3pt}
In this paper, we study the novel problem of integrating online and offline COVID-19-related data to seek common and distinct topics across them. Finding the commonness and distinctiveness between online and offline data is an important research problem because it provides an effective way for researchers to study how users' opinions change over time and how offline news sources affect online social media discussions. Furthermore, studying this problem paves the way for internet users to gather information on COVID-19-related topics and avoid possible misinformation effectively. To this end, we use the joint ONMF method to compute common and distinct topics efficiently. We design qualitative and quantitative experiments to measure the performance of joint ONMF and provide valuable insights into how offline data affects online data. For future work, we will study methods that can automate the process of finding interesting points of view and user attributes contributed to online discussions about offline events.

\section*{Acknowledgement}
\vspace{-0.1cm}
This work was supported by the Office of Naval Research under Award No. N00014-21-1-4002. Opinions, interpretations, conclusions, and recommendations are those of the authors.
\vspace{-0.1cm}

\bibliography{references}

\begin{thebibliography}{10}
\providecommand{\url}[1]{\texttt{#1}}
\providecommand{\urlprefix}{URL }
\providecommand{\doi}[1]{https://doi.org/#1}

\bibitem{cao2007detect}
Cao, B., et~al.: Detect and track latent factors with online nonnegative matrix
  factorization. In: IJCAI (2007)

\bibitem{cheng2020tracking}
Cheng, L., et~al.: Tracking disaster footprints with social streaming data. In:
  IAAI (2020)

\bibitem{cotfas2021longest}
Cotfas, L.A., et~al.: The longest month: analyzing covid-19 vaccination
  opinions dynamics from tweets in the month following the first vaccine
  announcement. IEEE Access  (2021)

\bibitem{crupi2022echoes}
Crupi, G., et~al.: Echoes through time: Evolution of the italian covid-19
  vaccination debate. In: ICWSM (2022)

\bibitem{feng2021integrating}
Feng, S., Kirkley, A.: Integrating online and offline data for crisis
  management: Online geolocalized emotion, policy response, and local mobility
  during the covid crisis. Scientific Reports  (2021)

\bibitem{glandt2021stance}
Glandt, K., et~al.: Stance detection in covid-19 tweets. In: IJCNLP (2021)

\bibitem{jiang2022covaxnet}
Jiang, B., Sheth, P., Li, B., Liu, H.: Covaxnet: An online-offline data
  repository for covid-19 vaccine hesitancy research. arXiv preprint
  arXiv:2207.01505  (2022)

\bibitem{kim2015simultaneous}
Kim, H., et~al.: Simultaneous discovery of common and discriminative topics via
  joint nonnegative matrix factorization. In: KDD (2015)

\bibitem{lee2000algorithms}
Lee, D., Seung, H.S.: Algorithms for non-negative matrix factorization. NeurIPS
   (2000)

\bibitem{lee1999learning}
Lee, D.D., et~al.: Learning the parts of objects by non-negative matrix
  factorization. Nature  (1999)

\bibitem{lwin2020global}
Lwin, M.O., et~al.: Global sentiments surrounding the covid-19 pandemic on
  twitter: analysis of twitter trends. JMIR public health and surveillance
  (2020)

\bibitem{lyu2022misinformation}
Lyu, H., et~al.: Misinformation versus facts: Understanding the influence of
  news regarding covid-19 vaccines on vaccine uptake. Health Data Science
  (2022)

\bibitem{macdonald2015vaccine}
MacDonald, N.E., et~al.: Vaccine hesitancy: Definition, scope and determinants.
  Vaccine  (2015)

\bibitem{martin2017leveraging}
Mart{\'\i}n, Y., et~al.: Leveraging twitter to gauge evacuation compliance:
  Spatiotemporal analysis of hurricane matthew. PLoS one  (2017)

\bibitem{pierri2022online}
Pierri, F., et~al.: Online misinformation is linked to early covid-19
  vaccination hesitancy and refusal. Scientific reports  (2022)

\bibitem{poddar2022winds}
Poddar, S., et~al.: Winds of change: Impact of covid-19 on vaccine-related
  opinions of twitter users. In: ICWSM (2022)

\bibitem{reynolds2002crisis}
Reynolds, B., et~al.: Crisis and emergency risk communication  (2002)

\bibitem{shen2020using}
Shen, C., et~al.: Using reports of symptoms and diagnoses on social media to
  predict covid-19 case counts in mainland china: Observational infoveillance
  study. JMIR  (2020)

\bibitem{wicke2021covid}
Wicke, P., Bolognesi, M.M.: Covid-19 discourse on twitter: How the topics,
  sentiments, subjectivity, and figurative frames changed over time. Frontiers
  in Communication  (2021)

\bibitem{wilson2020social}
Wilson, S.L., Wiysonge, C.: Social media and vaccine hesitancy. BMJ global
  health  (2020)

\bibitem{xiong2021social}
Xiong, Z., et~al.: Social media opinions on working from home in the united
  states during the covid-19 pandemic: Observational study. JMIR medical
  informatics  (2021)

\bibitem{yeo2020disaster}
Yeo, J., et~al.: Disaster recovery communication in the digital era: Social
  media and the 2016 southern louisiana flood. Risk analysis  (2020)

\end{thebibliography}
\bibliographystyle{splncs04}

\end{document}